\documentclass[1p]{Article_website}%%%%where rsos is the template name

\usepackage{graphicx}
\usepackage{cleveref}
\usepackage{cancel}
\usepackage[normalem]{ulem}
\usepackage{amsmath}
\usepackage{amssymb}
\usepackage{psfrag}
\usepackage{pstricks}
\usepackage{fancyheadings}
\usepackage{enumerate}
\usepackage{nicefrac}
\usepackage{graphicx}
\usepackage{gensymb}
\usepackage{textcomp}
\usepackage{subfigure}
\usepackage[misc,geometry]{ifsym} 
\usepackage{cleveref}
\usepackage{wasysym}
\usepackage{color,soul}
\usepackage[toc,page]{appendix}
\usepackage{times}
\usepackage[percent]{overpic}
\usepackage{color}
\usepackage{tikz}
\newcommand{\txb}[1]{\textcolor{black}{#1}}

\begin{document}

%%%% Article title to be placed here
\title{\txb{Experiments and modelling of rate-dependent transition delay in a stochastic subcritical bifurcation}}
%Rate-dependent critical transitions in stochastic bifurcations

\author{%%%% Author details
Giacomo Bonciolini$^1$, Dominik Ebi $^2$,  Edouard Boujo$^1$, Nicolas Noiray$^1$}

%%%%%%%%% Insert author address here
\address{$^{1}$ CAPS Laboratory, MAVT department ETH Z\"urich, Sonneggstrasse 3, 8092, Zurich, Switzerland\\
$^{2}$Laboratory for Thermal Processes and Combustion, Paul Scherrer Institute, 5232, Villigen, Switzerland}

%%%% Keyword entries to be placed here %%%%
%\keywords{Critical transitions, Bifurcation delay, Thermoacoustics, Stochastic systems}

%%%% Insert corresponding author and its email address}
%\corres{Giacomo Bonciolini\\
%\email{giacomob@ethz.ch}\\
%Nicolas Noiray\\
%\email{noirayn@ethz.ch}}

%%%% Abstract text to be placed here %%%%%%%%%%%%
\begin{abstract}
Complex systems exhibiting critical transitions when one of their governing parameters varies are ubiquitous in nature and in engineering applications. Despite a vast literature focusing on this topic, there are few studies dealing with the effect of the rate of change of the bifurcation parameter on the tipping points. In this work, we consider a subcritical stochastic Hopf bifurcation under two scenarios: the bifurcation parameter is first changed in a quasi-steady manner and then, with a finite ramping rate. In the latter case, a rate-dependent bifurcation delay is observed and exemplified experimentally using a thermoacoustic instability in a combustion chamber. 
\txb{This delay increases with the rate of change. This leads to a state transition of larger amplitude compared to the one that would be experienced by the system with a quasi-steady change of the parameter.}
%In short, the faster the ramping rate is, the later the tipping happens in the bifurcation diagram and the more intense the state transition is. 
We also bring experimental evidence of a dynamic hysteresis caused by the bifurcation delay when the parameter is ramped back. A surrogate model is derived in order to predict the statistic of these delays and to scrutinise the underlying stochastic dynamics. Our study highlights the dramatic influence of a finite rate of change of  bifurcation parameters upon tipping points and it pinpoints the crucial need of considering this effect when investigating critical transitions.
\end{abstract}
%%%%%%%%%%%%%%%%%%%%%%%%%%%

%%%%%%%%%% Insert the texts which can accomdate on firstpage in the tag "fmtext" %%%%%

\maketitle
%%%%%%%%%%%%%%% End of first page %%%%%%%%%%%%%%%%%%%%%

\section{Introduction}
Many systems exhibit abrupt changes, or tipping, e.g. population extinction \cite{drake2010early,dodorico2005noise}, emergence of infectious diseases \cite{dibble2016waiting}, \txb{financial systems crisis \cite{may2008complex}}, compression buckling of mechanical structures \cite{vella2009macroscopic}, and climate transitions \cite{lenton2008tipping,ditlevsen2010tipping,turney2017rapid}. 
Tipping is dangerous if some states of the system are associated with extreme or catastrophic events, \txb{and this explains the interest this subject has received in the last decades. Recently, different studies demonstrated that economical or environmental disasters can be modelled as dynamical systems incurring a tipping. Therefore, the development of tipping forecasting techniques with early indicators is an active research area \cite{scheffer2009early,kuehn2011mathematical,scheffer2012anticipating}}.\\
A key aspect in this context is the distinction between three types of tipping, rooted in different causes \cite{ashwin2012tipping}.\\
\textit{B-tipping} is induced by a \textit{Bifurcation} where the system state changes drastically for a small change of a control parameter. \txb{In this case, the tipping can be often predicted with  techniques that rely on the popular concept of critical slowing down \cite{nazarimehr2017can,dakos2008slowing,dakos2015resilience,lenton2012early,meisel2015critical}, or that make use of other properties of the attractor \cite{karnatak2017early,jiang2018predicting}}.\\
In \textit{N-tipping}, dynamic \textit{Noise} induces jumps between several coexisting attractors \txb{(e.g. \cite{sutera1981stochastic,sura2002noise,semenov2017noise,nikolaou2015detection}); in this case, the analysis of the time series statistic can help in detecting precursor of critical transitions \cite{carpenter2006rising,chen2018rising}}.\\
\textit{R-tipping} is induced by the \textit{Rate} at which a control parameter is varied\txb{, if several possible attractors are present in the range of parameter variation. Inertial effects play a central role in R-tipping. In the case of standard R-tipping, the system starts from an attractor but, if the parameter rate of change is larger than a critical value, it cannot follow this attractor and tips to another one. \cite{ashwin2017parameter,siteur2016ecosystems,chen2015patterned,wieczorek2011excitability}. In the case of ``preconditioned R-tipping'', the system starts from an unstable condition and, depending on the rate of change, it  evolves towards one of the possible attractors \cite{tony2017experimental}}.\\
\txb{Inertial effects can also delay the bifurcation, moving the tipping point to higher/lower values of the bifurcation parameter as observed, for example, by Baer and Gaekel in \cite{baer2008slow} for the FitzHugh-Nagumo model. This delay is in general a function of the parameter rate of change. Therefore we will refer to this effect as \emph{``Rate-delayed tipping''}}.\\

%%%%%%%%%%%%%%%%%%%%%%%%%%%%%%%
%%%%%%%%%%%%%%%%%%%%%%%%%%%%%%%%%%%%%%%%%%%%%%%FIGURE4
\begin{figure}[]
\centering
\includegraphics[width=0.75\columnwidth]{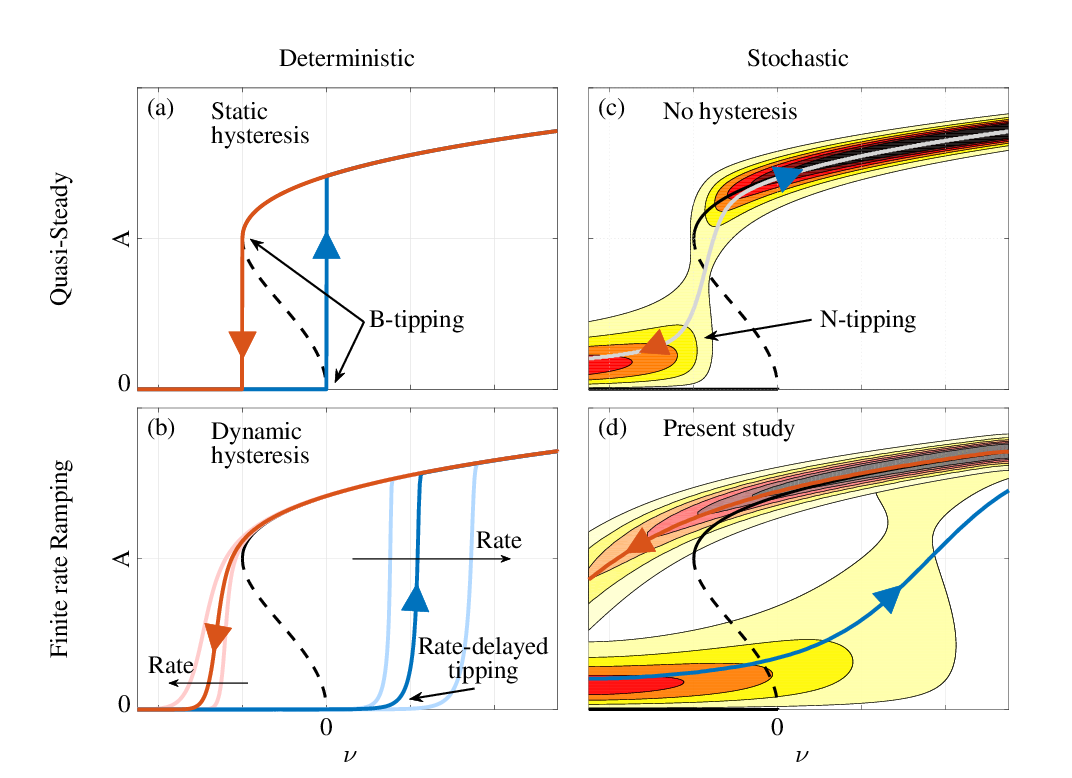}
\caption{Illustration of the various types of tipping encountered in the vicinity of the bi-stable region of a sub-critical bifurcation according to the classification proposed in \cite{ashwin2012tipping}. Solid and dashed black lines: deterministic attractor and repeller, respectively. Light to dark hues: low to high probability density. a) to c) B-tipping, \txb{Rate-delayed tipping} and N-tipping. d) Present work where B-, N- and \txb{Rate-delayed} tipping mechanisms occur simultaneously (see \cref{fig:03_ramp_exp} for the experimental data).}
\label{fig:04_summary}
\end{figure}
\txb{All those} mechanisms can manifest independently, or, like in the present study, simultaneously. 
In this case, the evolution of the system  results from the interplay of different time scales set by the ramp rate, the noise intensity and the system relaxation time \cite{shi2016towards}. \txb{Several examples can be found in the recent literature. Ashwin et al. \cite{ashwin2017fast} study the regimes of transition and the escape time in a network of bistable nodes as a function of the coupling and noise strengths. Sun and coworkers \cite{sun2015delay} assess the possibility of tipping for a Duffing-Van der Pol oscillator with time-delayed feedback, as a function of forcing noise intensity, feedback time delay and feedback intensity. The work from Clements and Ozgul \cite{clements2016rate} deals with two stochastic models for population dynamics, and studies the effect of the rate of change of one governing parameter on the system dynamics and on the predictability of tipping}. 
\txb{Berglund and Gentz \cite{berglund2002pathwise} provide theoretical and numerical analyses for rate-delayed tipping in presence of noise in supercritical pitchfork bifurcations. An analogous study is carried out by Ritchie and Sieber in \cite{ritchie2016early} for a rate-dependent tipping in a saddle-node bifurcation.}
%Berglund and Gentz \cite{berglund2002pathwise}, Ritchie and Sieber \cite{ritchie2016early} provide theoretical and numerical analyses \txb{respectively for rate-delayed and} rate-dependent tipping in presence of noise, respectively in supercritical pitchfork and saddle-node bifurcations. 
\txb{Kwasniok \cite{kwasniok2015forecasting} introduces a method to predict a fold and a Hopf bifurcation in presence of noise. Kuehn \cite{kuehn2017uncertainty} studies the delay in a Hopf bifurcation with a random initial condition}.\\ 

In this study, we show experimental evidence of simultaneous B-, N- and Rate-delayed tipping mechanisms at a Hopf subcritical bifurcation, in a lab-scale combustor subject to thermoacoustic instabilities in presence of turbulence-induced noise.\\ 
The four panels in figure \ref{fig:04_summary} illustrate how the three types of tipping combine in our system. In these diagrams, the amplitude $A$ is plotted as a function of the bifurcation parameter $\nu$.
In the absence of noise and for a quasi-steady change of the bifurcation parameter (figure \ref{fig:04_summary}a), the system state evolves on the deterministic attractor, leading to B-tipping and hysteresis (blue and red for increasing and decreasing $\nu$). This quasi-steady picture changes if the bifurcation parameter varies at a finite rate (figure \ref{fig:04_summary}b): bifurcation delay occurs, and it is a function of the rate (e.g. \cite{premraj2016experimental,holden1993slow,bergeot2014response}). For a quasi-steady variation of $\nu$ in the presence of stochastic forcing (\cref{fig:04_summary}c), the hysteresis is suppressed in a statistical sense. For each value of the bifurcation parameter, the state is defined by a probability density distribution. In this case, N-tipping occurs in the bistable region (e.g. \cite{samoilov2005stochastic,lenton2008tipping}). Finally, when the bifurcation parameter is varied at a finite rate in presence of stochastic forcing (\cref{fig:04_summary}d), the highest probability of state transition is delayed. This is the case discussed in this work. Our scenario therefore results from the combination of a finite-rate ramping through a stochastic subcritical bifurcation. \\
%%%%%%%%%%%%%%%%%%%%%%%%%%%%%%%%%%%%%%%%%%

\txb{This paper is organised as follows. In \cref{sec:TA_inst}, we introduce the physical problem of thermoacoustic instabilities. In \cref{ss:stat_exp} and \cref{ss:ramp_exp} we} show experimental results where the average tipping point is delayed when the control parameter is ramped at a finite rate. 
In \cref{ss:stat_mod} and \cref{ss:ramp_mod} we develop a low-order stochastic  model of the system and demonstrate with a quantitative first-passage time analysis how the bifurcation delay statistic varies with the ramping rate.
Finally, in \cref{s:FPA} we consider a  situation where a control parameter is  ramped up and, if tipping is detected, ramped down in order to come back to the initial safe state.
In this situation, the system may suffer irreversible damage if the ramp up is too fast, which applies to many industrial applications or to natural systems like, for instance, climate transitions. 
%%%%%%%%%%%%%%%%%%%%%%%%%%%%%%%%%%%%%%%%%%%%FIG1
\begin{figure*}[h!]
\centering
\includegraphics[trim= 0 0 0 0, clip, width=\textwidth]{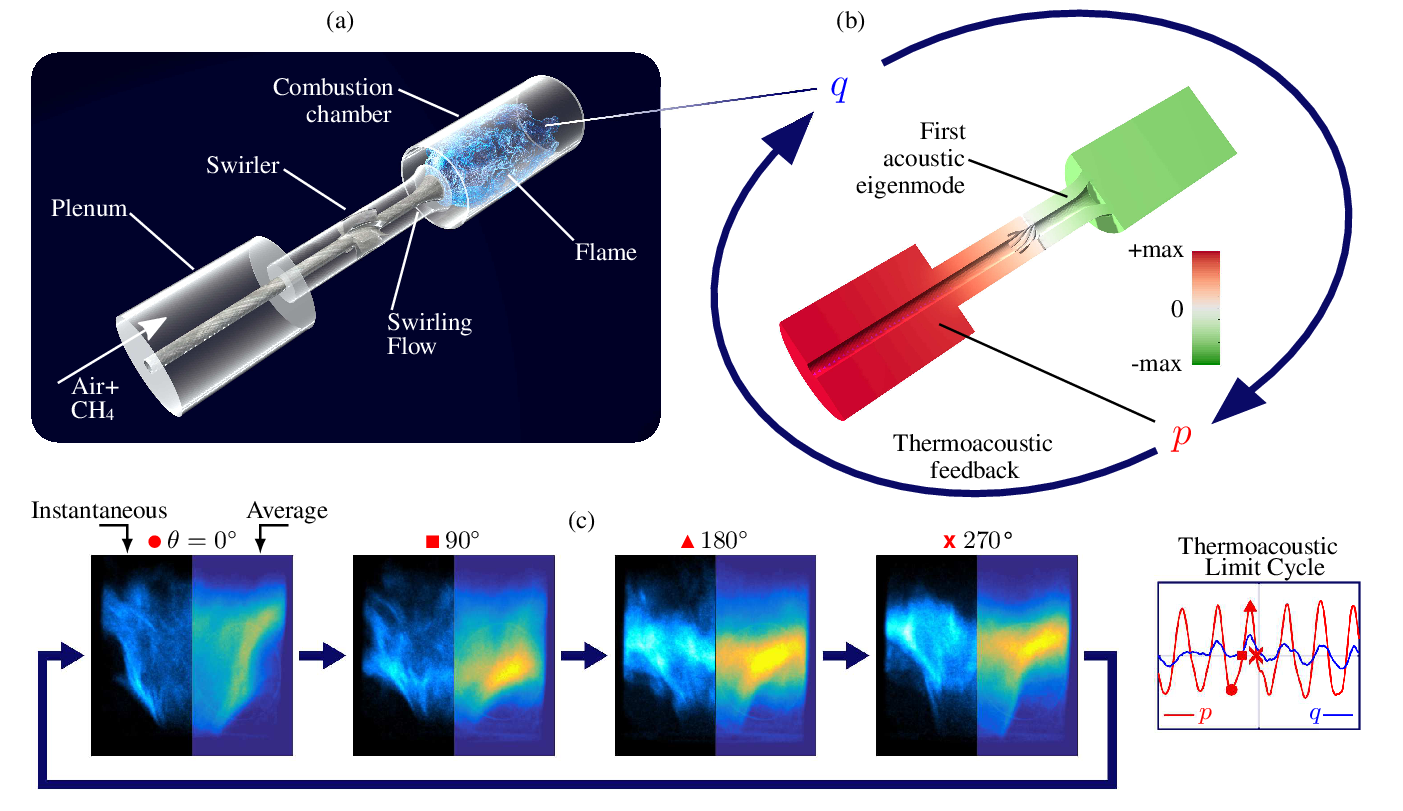}
\caption{Thermoacoustic instabilities occur in combustion chambers for aeronautic and power-generation applications. a) Schematic of our lab-scale swirled combustor. b) Illustration of this unstable  coupling. Under a certain phase difference relationship, well known as the Rayleigh criterion, a constructive feedback establishes between the unsteady heat release rate $q(\boldsymbol{x},t)$ and the acoustic field $p(\boldsymbol{x},t)$. c) Thermoacoustic limit cycle in the lab-scale combustor used for this work. In the left loop, four snapshots of the flame showing the coherent motion of the flame leading to $q$ and originating from the thermoacoustic feedback. Time traces of acoustic pressure and heat release rate are shown on the right.}  
\label{fig:01_TA}
\end{figure*}
%%%%%%%%%%%%%%%%%%%%%%

%%%%%%%%%%%%%%%%%%%%%%%%%%%%%%%%%%%%%%%%%%%%%%%%%%%%%%%%%%%%%%%%
\section{Thermoacoustic instabilities}
\label{sec:TA_inst}
Thermoacoustic coupling is a phenomenon that has fascinated scientists for over two centuries. In 1777, Dr. William Higgins reported, with surprise, on a hydrogen flame emitting ``sweet tones'' if placed inside a glass tube \cite{higgins1802}. In 1894, Lord Rayleigh provided an explanation to this observation: the gas in the tube resonates if the flame (or any other source) provides heat at the moment of maximum gas compression \cite{rayleigh1896}.\\
Many years after, during the Cold War, thermoacoustic instabilities became a very critical issue for one of the most challenging project ever realised by humankind: the Apollo program to take man to the Moon. As detailed in \cite{oefelein1993}, the F-1 engines propelling the Saturn V had destructive combustion instabilities that required 2000 full-scale tests, with empirical modifications of the chamber geometry before the rocket was ready for take off.\\
More recently, thermoacoustic instabilities became a recurrent issue in the development phase of heavy-duty gas turbines for power generation and   turbofans  for air transportation. This is because the resulting intense acoustic fields induce high-cycle fatigue of the combustion chambers \cite{lieuwen2012book}. For heavy-duty gas turbines, the pressing demand for machines with high power density and ultra low emissions, which are capable of compensating the production intermittency of the wind and solar sources, led to the  use of lean premixed flames. Unfortunately, these flames are more  prone to  thermoacoustic instabilities. In the case of airplane turbofans, these instabilities constitute an increasingly serious obstacle to the development of new aeroengines complying with more stringent emission regulations \cite{icao2016}.\\
The suppression of these instabilities is very challenging due to the uniqueness and complexity of engines in real life application \cite{poinsot2017}. 
Despite the achievements attained over the past decades in terms of passive mitigation implementation, development engineers cannot predict if a combustor prototype will have a sufficiently large pulsation-free operating window, over which the acoustic level is acceptable for the mechanical integrity of the components.

%
%%%%%%%%%%%%%%%%%%%%%%%%%%%%%%%%%%%%%%%%%%%%%%%%%%%%%%%%%
%%%%%%%%%%%%%%%%%%%%%%%%%%%%%%%%%%%%%%%%%%%%%%%%%%%%%%%%%
%
\Cref{fig:01_TA}a shows a schematic of our lab-scale combustor\footnote{Additional details about the combustor and the experimental apparatus are provided in the appendix.}. The air pre-mixed with methane enters the plenum, a volume that, in practice, evens the flow delivered by the compressor and guides it to the inlet of the burner. Then, the mixture passes through the swirler, a set of curved blades that rotate the flow. This rotational motion is essential to achieve a stable anchoring of the flame. Then, the flow enters the combustion chamber, where combustion takes place. 
At any operating point, the fluctuating component $q$ of the heat release rate $Q=\bar{Q}+q$ acts as a source term in the wave equation:
\begin{equation}
\label{eq:wave}
\dfrac{\partial^2 p}{\partial t^2} - c^2 \nabla^2 p = (\gamma-1)\dfrac{\partial q}{\partial t},
 %\left(\partial_{tt} - c^2\mathrm{\Delta}\right) p \,=\, (\gamma-1)\, \partial_t q,
\end{equation} 
where $p$ is the acoustic pressure, $c$ the speed of sound and $\gamma$ the specific heat ratio. In practice, the unsteady heat release of the flame $q$ is influenced by the acoustic field $p$, via, for instance, acoustically-triggered fuel supply modulation or coherent vortex shedding, which leads to a thermoacoustic feedback loop \cite{boujo2016quantifying}.
As illustrated in \cref{fig:01_TA}b, the geometry of the combustor and the temperature distribution define a set of acoustic modes in the chamber. 
Each mode is characterised by a shape and an eigenvalue. The latter determines whether the thermoacoustic mode is linearly stable or unstable. The system stability depends on several operating parameters, such as the mass flows of fuel $\dot{m}_\text{CH4}$ or air $\dot{m}_\text{air}$. The transition from linearly stable to linearly unstable regime occurs at Hopf bifurcations, where the sign \txb{of the growth rate of the mode} changes. If unstable, the thermoacoustic dynamics is characterised by a limit cycle, with amplitudes $p_\text{rms}$ and $q_\text{rms}$ being defined by the natural acoustic damping of the chamber, and by the linear and nonlinear components of the flame response to acoustic perturbations \cite{boujo2016quantifying,noiray2017method}. The non-coherent component of the heat release rate fluctuations, which is induced by turbulence, acts as a broadband forcing on this self-sustained thermoacoustic oscillation.\\
A typical operating condition for which we observe a thermoacoustic limit cycle is presented in \cref{fig:01_TA}c (see also the movie in the supplementary material). The four panels in the loop show instantaneous flame pictures and the corresponding phase-averaged flame shapes. The right plot displays the time traces of the acoustic pressure signal $p$ (in red) and the heat release rate $q$ (in blue) (note the symbols on the time trace corresponding to the four flame snapshot in the left loop). The flame exhibits a periodic motion at the frequency of the first acoustic mode (150~Hz), with sound intensity at the  anti-nodes exceeding 150 dB, which is considerable for a burner operated at atmospheric pressure. This dynamic state would not be acceptable in a commercial aeronautical engine or in a heavy-duty gas turbine, because the acoustic loading, which scales with the engine operating pressure, would be highly detrimental for the mechanical components.\\

In this work, we focus on the transient thermoacoustic dynamics associated with the passage through the Hopf bifurcation when one of the critical operating parameters -- the equivalence ratio -- is ramped. We show experimental evidence of a bifurcation delay and explain the phenomenon using a surrogate low-order model. This is particularly relevant for the development of new aeronautical and land-based gas turbines, which require fast loading or deloading, and which may be at risk due to such \txb{rate-delayed} tipping points.

%%%%%%%%%%%%%%%%%%%%%%%%%%%%%%%%%%%%%%%%%%%%%%%%%%%%
%%%%%%%%%%%%%%%%%%%%%%%%%%%%%%%%%%%%%%%%%%%%%%%%%%%
%%%%%%%%%%%%%%%%%%%%%%%%%%%%%%%%%%%%%%%%%%%%%%%%%%%
\section{Subcritical bifurcation}
\txb{This section presents two main results. In the first part, the results of the experimental mapping of the combustor dynamics as a function of the equivalence ratio are shown. In the second part, a low-order model of the system is derived.}
%%%%%%%%
%
%%%%%%%%%%%%%%%%%%%%%%%%%%%%%%%%%%%%%%%%%%%%%%%%FIGURE2
\begin{figure*}[h!]%[tbhp]
\centering
\includegraphics[width=\textwidth]{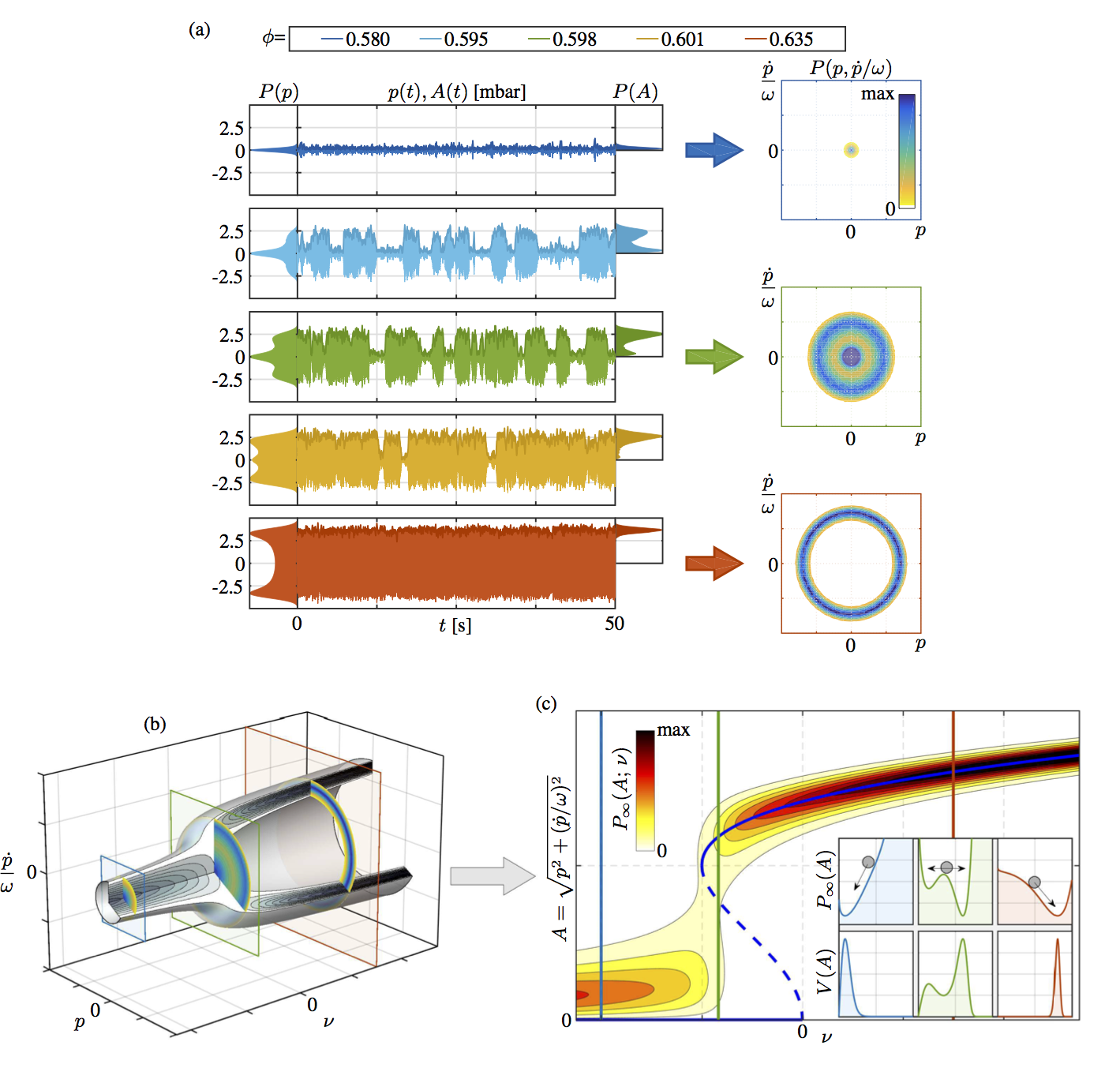}
\caption{a) Experimental records of the thermoacoustic subcritical Hopf bifurcation investigated in this work. According to the methane/air mixture equivalence ratio $\phi=(\dot{m}_\text{CH4}/\dot{m}_\text{air})/(\dot{m}_\text{CH4}/\dot{m}_\text{air})_\text{stoich.}$, the acoustic pressure recorded in the combustor has three different signatures, reflected in the different shapes of the PDFs $P(p)$ and $P(A)$. From top to bottom (increasing $\phi$): small amplitude acoustic pressure resulting from the forcing of the linearly stable thermoacoustic mode by turbulence-induced noise; bistable thermoacoustic dynamics with two intermittently visited  attractors; high amplitude limit cycle. These three possible regimes are also presented by the joint PDFs of the oscillation phase portrait $P(p,\dot{p}/\omega)$ at three exemplary $\phi$. b) Surrogate oscillator model \eqref{eq:oscillator} that mimics the subcritical Hopf bifurcation when the parameter $\nu$ is increased. In the 3D plot, $P_\infty(p,\dot{p}/\omega\, ; \nu)$ and three cuts, resembling the experimental $P(p,\dot{p}/\omega)$. c) The stationary PDF $P_\infty(A\,;\,\nu)$ for the slow-varying oscillation amplitude $A$, obtained with the transformation of variables $A^2=p^2+(\dot{p}/\omega)^2$. On top of it, the deterministic pitchfork and saddle-node bifurcation diagram (in blue), and the stationary PDF $P_\infty(A)$ with the corresponding potential $V(A)$  for an overdamped particle at the three selected $\nu$.}
\label{fig:02_bif}
\end{figure*}
%%%%%%%%%%%%%%%%%%%%%%%%%%%%%%%%%%%%%%%%%%%%%%%%%%%%
\subsection{Stationary experiment}
\label{ss:stat_exp}
The combustor was operated selecting one equivalence ratio $\phi$ at a time. The stationary operation was reached and a long acoustic pressure signal $p(t)$ was recorded using a microphone placed inside the chamber. The oscillation amplitude $A(t)$ was then extracted by applying the Hilbert transform to $p(t)$. The procedure was repeated for different equivalence ratios $\phi$ in the range [0.580; 0.635].  
The results for five selected $\phi$ are presented in \cref{fig:02_bif}a. On the left, the measured acoustic pressure and amplitude signals are plotted,
together with their probability density functions (PDFs) $P(p)$ and $P(A)$. On the right, the joint PDFs $P(p,\dot{p}/\omega)$ for three of the presented operative points show the statistic of the phase portraits.\\
These results demonstrate how the system undergoes a subcritical Hopf bifurcation when the control parameter is varied.  For low equivalence ratio $\phi$, the system state is attracted towards zero. The small fluctuations of the acoustic signal envelope are due to the stochastic forcing exerted by the intense turbulence in the combustor. For intermediate values of $\phi$, two states are possible: small amplitude acoustic pressure and high amplitude limit cycle. The intermittency between the two states is triggered by the turbulence-induced stochastic forcing (N-tipping, as in \cref{fig:04_summary}c). For higher equivalence ratio $\phi$, the stochastically-forced limit cycle is the only stable state. The reader can refer to the supplementary material for the movies of the three regimes.\\
%%%%%%%%%%%
\subsection{Non-linear oscillator model}
\label{ss:stat_mod}
\txb{
The thermoacoustic behaviour described in the previous section can be reproduced by a low-order model derived from first principles. The Helmholtz equation (\ref{eq:wave}) (hereafter rewritten in Laplace space) defines the acoustic pressure field in the combustor, given an unsteady source of heat in the volume and impedance conditions at the boundaries::
\begin{equation}
\label{helmholtz}
\nabla^{2}\widehat{p}(s,x)-\left(\frac{s}{c}\right)^2\widehat{p}(s,x)=-s\frac{(\gamma-1)}{c^{2}}\widehat{Q}(s,x) \,\, \text{in the domain},
\end{equation}	
\begin{equation}
\label{helmholtz_BC}
\frac{\widehat{p}(s,x)}{\mathbf{\widehat{u}}(s,x)\cdot \mathbf{n}}=Z(s,x) \,\,\,\,\, \text{on the boundaries},
\end{equation}
where $s$ is the Laplace variable, $\widehat{p}$ and $\mathbf{\widehat{u}}$ are the transforms of acoustic pressure and velocity fluctuations, $x$ the position, $c$ the local speed of sound, $\gamma$ the specific heat ratio, $\widehat{Q}$ the heat release rate source term, $\mathbf{n}$ the outward normal to the boundary and $Z$ the acoustic impedance. This equation is valid under low Mach number conditions.\\
Although non-linear coupling among different thermoacoustic modes can occur in some practical configurations, we focus on situations where, like in the present case, one mode is dominant in the thermoacoustic dynamics. Therefore, it is possible to project the acoustic field on an orthogonal basis $\mathbf{\Psi}$ and approximate the system's dynamics with the one of the dominant mode only, which will be denoted by $\psi$ \cite{lieuwen2003statistical,culick2006unsteady}. This yields the approximation $\widehat{p}(s,x)\approx\widehat{\eta}(s)\psi(x)$, $\widehat{\eta}$ being the mode amplitude:
\begin{equation}
\label{eq:mode}
\widehat{\eta}=\dfrac{s\rho c^2}{s^2+\omega^2}\dfrac{1}{V\Lambda}\left(\dfrac{\gamma-1}{\rho c^2}\int_V\widehat{Q}(s,x)\psi^*(x)\text{d}V-\widehat{\eta}\int_{\partial V}\dfrac{|\psi(x)|^2}{Z(s,x)}\text{d}\sigma\right),
\end{equation}
where $\rho$ is the gas density and $\Lambda$ the mode normalisation coefficient. This equation can be rewritten as:
\begin{equation}
\label{eq:mode_osc}
(s^2+\alpha s+\omega_0^2)\widehat{\eta}=s\widehat{q},
\end{equation}
\begin{equation}
\label{eq:mode_alpha}
\text{with} \quad \alpha=  \dfrac{\rho c^2}{V\Lambda}\int_{\partial V}\dfrac{|\psi(x)|^2}{Z(s,x)}\text{d}\sigma,
\end{equation}
\begin{equation}
\label{eq:mode_q}
\text{and} \quad \widehat{q}= \dfrac{\gamma-1}{V\Lambda}\int_V\widehat{Q}(s,x)\psi^*(x)\text{d}V.
\end{equation}
Therefore, the system dynamics can be approximated by a forced damped harmonic oscillator \eqref{eq:mode_osc} of resonance frequency $\omega_0$. The term $\alpha>0$ represents the damping mechanisms, and it is assumed to be constant, since the impedance at the boundaries is generally a smooth function of $s$ and therefore is not expected to vary significantly around $\omega_0$. The term $\widehat{q}$ is the result of the weighting on the mode shape of the volumetric heat release rate and can be decomposed into two contributions: $\widehat{q}=\widehat{q}_n+\widehat{q}_c$. The first component $\widehat{q}_n$ represents the non-coherent part of the heat release rate oscillations, induced by the intense turbulence present in practical combustors.  The term $\widehat{q}_c$ refers to the coherent heat release rate fluctuations, which result from a feedback interaction with the acoustic field established in the combustor. Hence, this term can be expressed as a non-linear function of the modal amplitude $\eta$. It is customary to simplify this function by replacing it with its truncated Taylor expansion \cite{lieuwen2003statistical,culick2006unsteady}. The linear term coefficient $\beta$ of this expansion defines, together with the linear damping $\alpha$, the linear stability of the system. Absorbing in the constants the mode shape $\psi(x_p)$ at the pressure probe location $x_p$ and considering only the odd terms of the series expansion up to the fifth order leads to the following oscillator model for the thermoacoustic system: 
}
\begin{equation}
\label{eq:oscillator}
\ddot{p}+\omega_0^2p=[2\nu +\kappa p^2-\mu p^4]\dot{p}+\xi(t),
\end{equation}
where $\nu=(\beta-\alpha)/2$ is the oscillation linear growth rate in $\mathrm{rad.s^{-1}}$ and $\kappa$ and $\mu$ the two positive constants that define the non-linear component of the oscillator response. The term $\xi(t)$ is a white noise forcing of intensity $\Gamma$ that models the non-coherent turbulence-induced heat release rate fluctuations.  
In \cref{fig:02_bif}b, the plot shows the stochastic Hopf bifurcation featured by this oscillator, as a function of the bifurcation parameter $\nu$. 
This three-dimensional representation of the stationary joint-probability density $P_\infty(p, \dot{p}/\omega\, ; \nu)$ is depicted together with 3 orthogonal cuts resembling the ones obtained from the experiments and showing that the bifurcation parameter $\nu$ of the surrogate model \eqref{eq:oscillator} corresponds to the equivalence ratio $\phi$ in the experiments.\\

It is convenient to describe the system evolution in terms of the slowly varying amplitude $A$ and phase $\varphi$, with $p(t)=A(t)\cos(\omega_0t+\varphi(t))$. By inserting this ansatz for $p$ into the second order stochastic differential equation \eqref{eq:oscillator} and by performing deterministic and stochastic averaging (e.g. \cite{noiray2017linear}), one gets first order Langevin equations for $A$ and $\varphi$. The equation for $A$ is $\mathrm{d}{A}/\mathrm{dt}=\mathcal{F}(A)+\zeta$, where $\mathcal{F}(A)=A\left[\nu+(\kappa/8)A^{2}-(\mu/16)A^4\right]+\Gamma/(4\omega_{0}^{2}A)$ and $\zeta$ is a white noise forcing of intensity $\Gamma/2\omega_0^2$. The deterministic dynamics derives from a potential with $\mathcal{F}(A)=-\mathrm{d}V/\mathrm{d}A$, and the equation does not depend on $\varphi$, which leads to the corresponding Fokker-Planck equation (FPE) for the variation in time of the amplitude PDF $P(A;t)$:
\begin{equation}
\label{eq:fp}
%\dfrac{\partial}{\partial t}P(A;t)=-\dfrac{\partial}{\partial A}[\mathcal{F}(A)P(A;t)]+\dfrac{\Gamma}{4\omega_0^2}\dfrac{\partial^2}{\partial A^2}P(A;t),
\dfrac{\partial P}{\partial t}=-\dfrac{\partial}{\partial A}[\mathcal{F}(A)P]+\dfrac{\Gamma}{4\omega_0^2}\dfrac{\partial^2P}{\partial A^2}
\end{equation}
Setting ${\partial P}/{\partial t}=0$, one obtains the stationary PDF $P_\infty(A\,;\,\nu)$, plotted in \cref{fig:02_bif}c as a function of the linear growth rate $\nu$, in a pitchfork bifurcation diagram fashion. To provide a visual reference, the bifurcation diagram of the deterministic case (i.e. in absence of noise, $\Gamma=0$) is superimposed in blue. This diagram shows the subcritical pitchfork and the saddle-node bifurcations governing the system.
In the bottom insets, the PDFs $P_\infty(A\,;\,\nu_i)$ for three selected values of the bifurcation parameter $\nu$ are presented.  In the upper insets, the corresponding potentials are plotted. The linearly stable and stable limit cycle conditions feature a single potential well at low or high amplitude, while the bistable case has two potential wells. 
The stochastic forcing causes the jumps from one basin of attraction to the other, and hence the intermittency  between low-amplitude noisy fluctuations and high-amplitude limit cycle oscillations.
%%%%%%%%%%%%%%%%%%%%%%%%%%%%%%%%%%%%%%%%%%%%%%%%%%%
%%%%%%%%%%%%%%%%%%%%%%%%%%%%%%%%%%%%%%%%%%%%%%%%%%%
%%%%%%%%%%%%%%%%%%%%%%%%%%%%%%%%%%%%%%%%%%%%%%%%%%%
\section{Ramping}
\txb{In this section, the dynamics of the system under transient operation is analysed. In the first part, experimental results obtained by ramping the bifurcation parameter are provided. They highlight the presence in the system dynamics of B- and N-tipping mechanisms combined with inertial and hysteresis effects. In the second part, the model introduced in \cref{ss:stat_mod} is used to study the influence of the ramp rate on the system dynamics.}
%%%%%%%%%%%%%%%%%%%%%%%%%%%%%%%%%%%%%%%%%%%%%%%FIGURE3
\begin{figure*}[h!]%[tbhp]
\centering
\includegraphics[width=\textwidth]{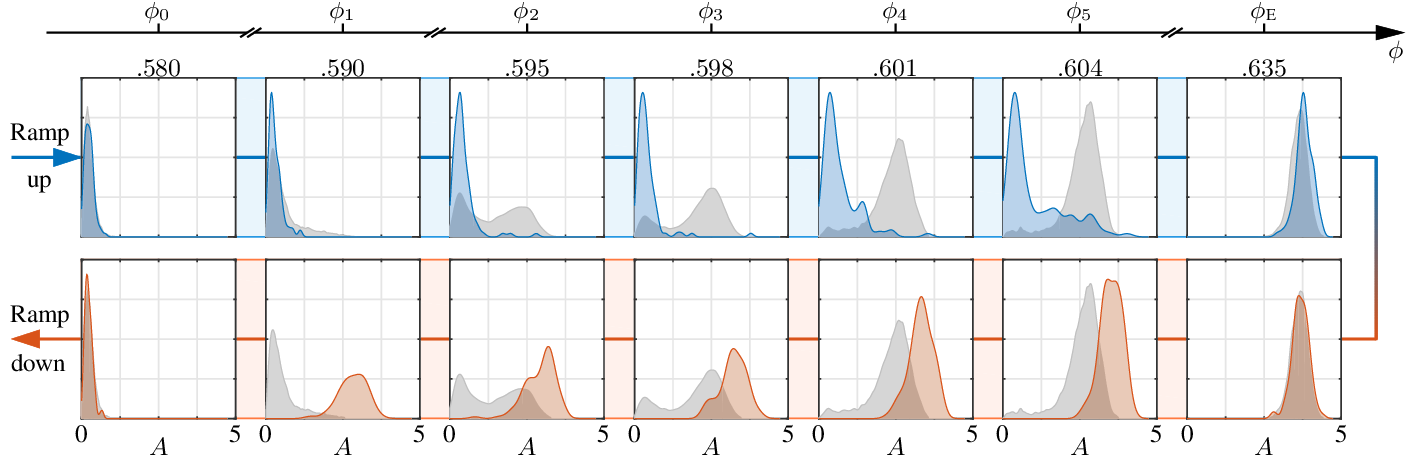}
\caption{Experimental evidence of the bifurcation delay and of the dynamic hysteresis in the ramped $\phi$ experiment. The panels are divided in ramp up (top row) and ramp down (bottom row). The  stationary probability density function $P_\infty(A;\,\phi_i)$ at seven equivalence ratios $\phi$ (grey) is given as a reference and compared to the evolution in time of the ensemble PDF $P(A;\,t_i)$, where $t_i$ is the time at which $\phi(t_i)=\phi_i$ with $\phi(t)=\phi_0+Rt$ for the ramp up and $\phi(t)=\phi_\text{E}-Rt$ for the ramp down.}
\label{fig:03_ramp_exp}
\end{figure*}
\subsection{Ramp experiment}
\label{ss:ramp_exp}
\txb{The following} test was performed on the test rig to highlight the peculiar dynamics of this combustor. The methane and air mass flows were controlled such that the equivalence ratio $\phi$ repeated 100 times the following four-step cycle: 1) linear increase for 4s from $\phi_0$ = 0.580 to $\phi_\text{E}$ = 0.635; 2) idle for 10s at $\phi_\text{E}$; 3) linear decrease for 4s back to $\phi_0$; 4) idle for 10s at the lowest equivalence ratio. \Cref{fig:03_ramp_exp} presents the results of this experiment. The panels are grouped in two rows: the upper row corresponds to the statistic of the 100 ramps up, the lower row to the one of the 100 ramps down. Each column corresponds to an equivalence ratio. The PDFs of this ramp experiment were obtained via a kernel density estimation (KDE) applied over the 100 realisations, and they are plotted in color (blue for the ramp up, red for the ramp down). In all the panels, the stationary PDF for the corresponding $\phi$ (no ramping, already presented in \cref{fig:02_bif}a) is given in grey as a reference.\\
The system experiences dynamic hysteresis: in the bistable region, even though the stationary PDF features two maxima, the system stays in the low-amplitude (resp. high-amplitude) range when $\phi$ is ramped up (resp. down). Another feature is the delay in transition, easily observable in the bottom row: the dynamic PDF peak is at an amplitude that is higher than the one of the stationary PDF at the same $\phi$. This means that the system experiences inertial effects, remaining close to the initial state longer: a bifurcation delay is observed. This observation corresponds to the case depicted in \cref{fig:04_summary}d.
%%%%%%%%%%%%%%%%%%%%%%%%%%%%%%%%%%%%%%%%%%%%%FIGURE5
\begin{figure}[h!]%[tbhp]
\centering
\includegraphics[width=0.75\columnwidth]{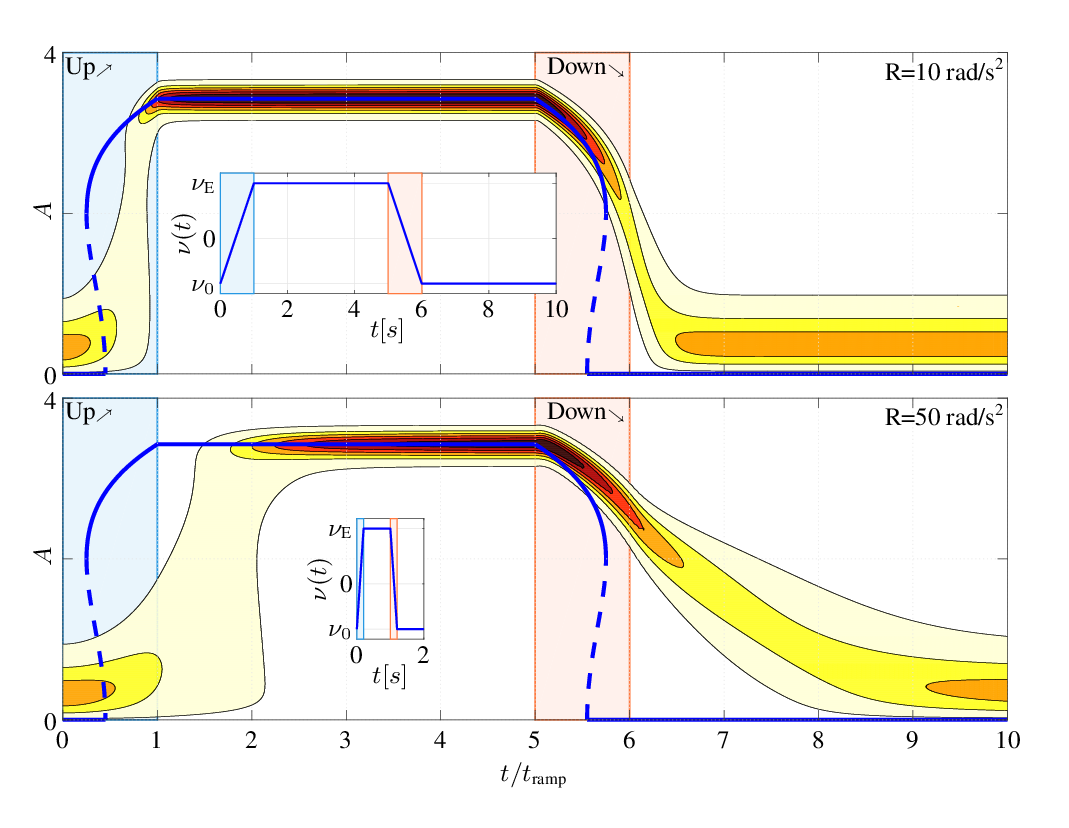}
\caption{Ramping cycle of $\nu$ between the values $\nu_0=-4.5$ and $\nu_\text{E}=5.5$ for the oscillator model \eqref{eq:oscillator} for two different ramp rates $R$. The contour plot represents the PDF $P(A;t)$ computed by time-marching the FPE \eqref{eq:fp}. The time is normalised with the ramp time $t_\text{ramp}=(\nu_\text{E}-\nu_0)/R$. In the insets, the ramp cycle in dimensional time. The other parameters of the model are: $\omega_0/2\pi=120$s\textsuperscript{-1}, $\kappa=8$s\textsuperscript{-1}, $\mu=2$s\textsuperscript{-1}, $\Gamma/4\omega_0^2=0.44$.}
\label{fig:05_ramp_sim}
\end{figure}
\\
\subsection{Rate-dependent bifurcation delay}
\label{ss:ramp_mod}
The ramp rate, together with the ramp profile, is expected to influence the bifurcation delay, as shown in \cite{baer2008slow} for a deterministic system. We therefore used the surrogate oscillator model to investigate this aspect in more detail. The parameter $\nu$ was varied linearly in time between two values $\nu_0$ and $\nu_\text{E}$ at different rates $R$. 
Two approaches were used. The first is to simulate \eqref{eq:oscillator} in Simulink, varying the initial condition and running different realisations of the process. Extracting the envelope for each realisation and considering the ensemble statistic, it is possible to draw maps of the evolution in time of the amplitude PDF $P(A;t)$. The other approach is to integrate numerically the FPE \eqref{eq:fp} and obtain directly $P(A;t)$. The two methods closely agree, as shown in the appendix. In \cref{fig:05_ramp_sim} the results of the FPE integration are presented. A ramp up/idle/ramp down/idle cycle is solved, for two different ramp rates $R=$50 rad/s\textsuperscript{2} and $R=$10 rad/s\textsuperscript{2}. The dynamic stochastic bifurcation delay is captured and it is observed that a faster ramp leads to a more pronounced delayed transition from one stable point to the other.\\ 
An important aspect of the phenomenon depicted in this figure is that the state transitions are delayed with respect to the bifurcation point, but not time delayed (the horizontal axis in these figures is normalised by the physical duration of the cycle). In other words,
a higher rate of change of the time-varying potential induces a faster transition into the neighbouring basin of attraction, but the transition occurs for a delayed value of the bifurcation parameter compared to the quasi-steady picture of the system. 
%%%%%%%%%%%
%%%%%%%%%%%%%%%%%%%%%%%%%%%%%%%%%%%%%%%%%%%%%%%%%%%%%%%%%%%%%%%
%%%%%%%%%%%%%%%%%%%%%%%%%%%%%%%%%%%%%%%%%%%%%%%%%%%%%%%%%%%%%%%
%%%%%%%%%%%%%%%%%%%%%%%%%%%%%%%%%%%%%%%%%%%%%%%%%%%%%%%%%%%%%%%
%
%%%%%%%%%%%%%%%%%%%%%%%%%%%%%%%%%%%%%%%%%FIGURE 6
\begin{figure}[h!]
\centering
\includegraphics[trim= 0 0 0 0, clip, width=0.5\textwidth]{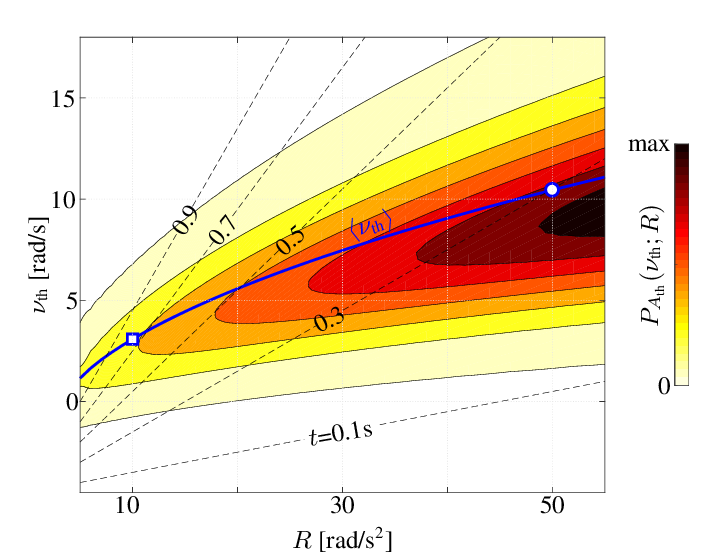}
\caption{$P_{A_\text{th}}(\nu_\text{th};\,R)$ is the probability density of the instantaneous linear growth rate $\nu$ at the first passage over the threshold amplitude $A_\text{th}$, as function of the ramp rate $R$. It is obtained from simulations of the unsteady FPE with absorbing boundary at $A=A_\text{th}$. In blue $\langle\nu_\text{th}(R)\rangle$, the linear growth rate of the system at the mean first passage time.} 
\label{fig:06_fpt}
\end{figure}

%%%%%%%%%%%%%%%%%%%%%%%%%%%%%%%%%%%%%%%%FIGURE 7
\begin{figure}[h!]
\centering
\includegraphics[width=0.75\textwidth]{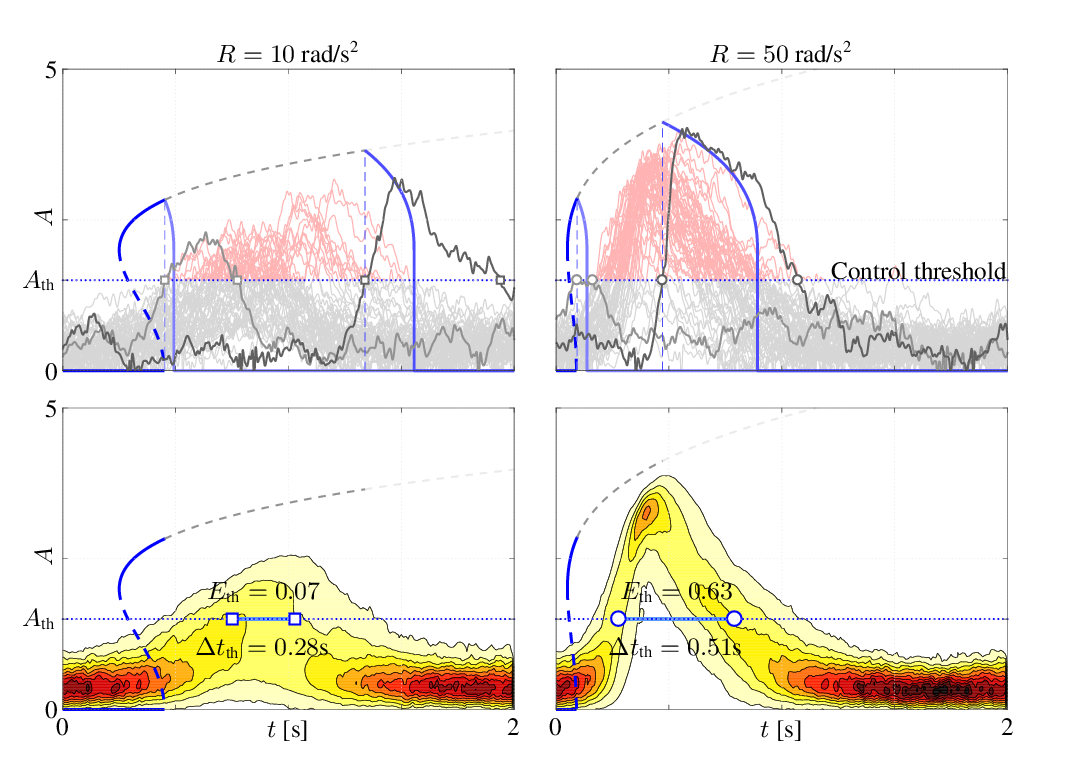}
\caption{Two exemplary cases (square $R$=10, circle $R$=50) are simulated in Simulink, implementing a control system that ramps down $\nu$ if the danger level is reached. 
Top row: different realisations of the process (thin lines, grey and red in the safety and danger zones, respectively)
and the two extreme realisations in terms of first passage time (thick grey lines) with their associated quasi-steady deterministic bifurcation diagrams (blue lines). 
Bottom row: KDE of the PDFs. The mean residence and mean released energy in the danger zone are indicated too.
}
\label{fig:07_control}
\end{figure}
%%%%%%%%%%%%%%%%%%%%%%%%%%%%%%%%%%%%%%%%%%%%%%%%%%%%%%

%%%%%%%%%%%%%%%%%%%%%%%%%%%%%%%%%%%%%%%%%%%%%%%%%%%%%%
%%%%%%%%%%%%%%%%%%%%%%%%%%%%%%%%%%%%%%%%%%%%%%%%%%%%%%
%%%%%%%%%%%%%%%%%%%%%%%%%%%%%%%%%%%%%%%%%%%%%%%%%%%%%%
\section{First Passage Analysis}
\label{s:FPA}
In this section, we imagine that a tipping point is feared due to the monotonous change of a key parameter of the system, and that one wants to ramp back this parameter sufficiently early to avoid the critical transition. In that situation, the underlying time-varying potential landscape is unknown and a controller monitors the state of the system while the parameter varies. 
As is usually done when new prototype engines are tested, we will take the current state of the system to feed the controller. Indeed, gas turbines and aeronautical engine combustors are equipped with a controller that constantly monitors the acoustic pressure level in the chamber. In case the measured acoustic pressure is too high, the control system intervenes, either changing the parameters to bring the operating condition back to a safe point, or in extreme cases, shutting off the flame by closing the fuel supply valve.\\
If the combustor features a subcritical bifurcation on the varied parameter (e.g. $\phi$), the system inertia is a factor that has to be taken into account. In this case, the transition from low to high amplitudes happens suddenly and, if the bifurcation delay is long, the reached acoustic pressure level can be considerable. In this situation, the control system detects the danger late and might be ineffective in avoiding damages to the system. 
A way to estimate the hazard represented by the delayed bifurcation is to compute, using the surrogate model, the statistic of the time $t_\text{fp}$ needed to reach a certain danger level. This is similar to the classical problem of first passage time, often addressed in the context of bifurcation theory for stochastic dynamics in steady double-well  potential \cite{torrent1988stochastic,kuske1999probability,miller2012escape,hu2010first,dibble2016waiting}. A major difference in the present situation is that the potential evolves with time. Ramp rate and noise intensity are expected to influence this escape problem as theoretically shown for other types of bifurcation in \cite{ritchie2017probability} or \cite{berglund2002beyond}. The statistic of the first passage time can be computed either performing an ensemble average over many time-domain simulations of the process, or solving the unsteady Fokker-Planck equation and imposing an absorbing boundary condition at that threshold level. Details about the two methods, with results in close agreement, are provided in the appendix.
The value $\nu(t_\text{th})=\nu_\text{th}$ of the control parameter $\nu(t)$ at the first passage time $t_\text{th}$ is of particular interest: this quantity is proportional to the danger of the delayed transition, as it determines the limit cycle amplitude when the transition occurs. This $\nu_\text{th}$ statistic can be determined as $\nu_\text{th}= \nu_0+Rt_\text{th}$. 
The results are presented in \cref{fig:06_fpt}. The contour levels represent the  probability density of $\nu_{th}$ as a function of the ramp rate $R$.  The mean value of $\nu_{th}$ (plotted in blue, $\langle\nu_{th}\rangle$) increases with the ramp rate $R$, while the time needed to reach the danger level is shorter (see the iso-time lines).  This finding indicates that a fast ramp of the control parameter is dangerous if a subcritical bifurcation is present, as exemplified in the two test cases presented in \cref{fig:07_control}. Here the process was simulated in Simulink: the parameter $\nu$ was ramped up at two different rates $R$ (10 rad/s\textsuperscript{2} and 50 rad/s\textsuperscript{2}) and when the danger level was reached, ramped back down at the maximum rate $R$=-50 rad/s\textsuperscript{2}.  In the top row, many realisations of this process are presented. As a function of the initial condition and of the random excitation, each realisation has a different evolution and, therefore, a different first passage time. 
The two extreme realisations (shortest and longest first passage times) are highlighted with thick lines. The respective deterministic bifurcation diagrams are superimposed to provide a visual reference. The PDFs obtained with a KDE over the realisations are plotted in the bottom row. The control system effectively brings the oscillations back to a safe level in both cases. However, the combined action of the finite ramp-down rate, dynamic hysteresis and inertia causes the system to stay in the danger zone for a certain time. The faster case $R$=50 rad/s\textsuperscript{2} is more critical: as discussed before, the crossing of the threshold level happens on average when the target $\nu$ is already high. As a result, the system abruptly reaches high-amplitude oscillations and has to travel a long distance on the bifurcation diagram upper branch before reaching the safety zone.  
This effect can be gauged by comparing two quantities for the two cases $R=10$ and $50$~rad/s\textsuperscript{2}: in the latter case, the mean residence time over the safety threshold $\Delta t_\text{th}$ is twice larger and the mean released energy $\langle E_\text{th}\rangle \propto (1/ \Delta t_{th})\int_{\Delta t_\text{th}}A^2\,\mathrm{d}t$ is nine times larger.
%
%%%%%%%%%%%%%%%%%%%%%%%%%%%%%%%%%%%%%%%%%%%%%%%%%%%%%%
%%%%%%%%%%%%%%%%%%%%%%%%%%%%%%%%%%%%%%%%%%%%%%%%%%%%%%
\section{Conclusions}
A subcritical Hopf bifurcation of a thermoacoustic system was investigated in this work.  A lab-scale combustor was operated under different values of  methane/air equivalence ratio, which serves as bifurcation parameter: depending on its value, acoustic pressure amplitude in the chamber is either damped, intermittently switching between low and high amplitudes, or attracted towards high-amplitude, which corresponds to a stable limit cycle. The main focus of the work was on the transient dynamics: the equivalence ratio was ramped in time and dynamic hysteresis and delayed bifurcation were observed.  
A non-linear oscillator surrogate model was used to investigate the effect of the ramp rate on the bifurcation delay. It was shown that when the control parameter is ramped faster, the transition from the damped regime to the limit cycle occurs for higher values of the bifurcation parameter. The corresponding first passage problem in a time-varying potential was solved with the unsteady Fokker-Planck equation and with Monte Carlo simulations of the process. 
This study primarily addresses a major problem of practical combustion systems. Operating conditions of gas turbines are often varied in time, for matching power grid requirement, and similar rapid changes of the combustion regimes also occur in aeronautical engines, especially at take-off. If a subcritical thermoacoustic bifurcation is present, a delayed bifurcation results in a sudden and unexpected acoustic pressure level rise, which is detrimental to the machine integrity. Therefore a slow variation of the machine parameters is advisable, especially when mapping the operating points of a new combustor. More broadly, this study is relevant for the countless systems, which exhibit critical transitions. This work highlights the importance of carefully considering the rate of change of the bifurcation parameter, when investigating tipping points.\\

\emph{Data accessibility.} The datasets supporting this article are available at doi:10.5061/dryad.4cj4k.\\
\emph{Author contributions.} N.N., G.B. and E.B. designed research; G.B. performed the numerical simulations and analysed all the data; D.E. made the experiments and supported in the analysis of experimental data; N.N. and E.B. provided scientific advises and helped in analysing data. G.B. and N.N. wrote the paper. All the authors proofread and made suggestions about the manuscript.\\
\emph{Competing interests.} We declare we have no competing interests.\\
\emph{Funding.} This research is supported by the Swiss National Science Foundation under Grant 160579.

% Bibliography
\bibliographystyle{apsrev_mod} %%%% .BST file
\bibliography{RSOS_cited_papers}

%
%%%%%%%%%%%%%%%%%%%%%%%%%%%%%%%%%%%%%
%%%%%%%%%%%%%%%%%%%%%%%%%%%%%%%%%%%%%%
%%%%%%%%%%%%%%%%%%%%%%%%%%%%%%%%%%%%%%
%
%%%%%%%%%%%%%%%%%%%%%%%%%%%%%%%%%%%%%%%%APPENDIX
\section*{Appendix}
%
%%%%%%%%%%%%%%%%%%%%%%%%%%%%%%%%%%%%%%%%%%%%%%%%%%%%%%%%%%%%%%%%%%%%%%%%%%%%%%%%%%%
\subsection*{Experimental setup}
\label{SUP_experiments}
The experiments were conducted using a turbulent, small-scale, swirled  combustor operated at atmospheric pressure. 
Electrically heated air ($300~^{\circ}$C) and methane are premixed upstream of the plenum.
The mixture then goes through a swirler and reaches the combustion chamber. 
The total mass flow rate was kept constant, with a bulk velocity of 10~m/s at the combustor inlet. 
The equivalence ratio was varied from $\phi=0.580$ to $\phi=0.635$, corresponding to a thermal power of about 12 kW.  
The inner and outer diameters of the eight-blade axial swirler are 19~mm and 41~mm respectively. 
This swirler imparts rotation to the flow, with a swirl number of 0.6. 
A quartz window located on one side of the cylindrical combustion chamber provides optical access to the flame. 
\\
Local acoustic pressure and spatially integrated OH$^*$ chemiluminescence were  acquired synchronously at a rate of 10~kHz. 
Acoustic pressure was recorded by means of four calibrated water-cooled microphones (Br\"uel\&Kj{\ae}r, type 4939) at several axial locations  (-235~mm, 25~mm, 115~mm and 245~mm from the combustion chamber inlet). 
The OH$^*$ chemiluminescence intensity was recorded using a photomultiplier equipped with a 310~nm bandpass-filter. For such perfectly premixed lean flames, this signal can be considered as proportional to the heat release rate.\\ 
The high-speed movies (1000 fps) of the turbulent flame are obtained with a LaVision HSS X camera coupled to a HS-IRO image intensifier. 
The  UV-optimised lens (Nikkor 105~mm $f/4.5$) of the intensifier is equipped with a 310~nm filter, which band-passes the OH$^*$ chemiluminescence.
%
%%%%%%%%%%%%%%%%%%%%%%%%%%%%%%%%%%%%%%%%%%%%%%%%%%%%%%%%%%%%%%%%%%%%%%%%%%
\subsection*{Ramping of the growth rate $\nu$: validation of the FPE method}
\label{SUP_validation_FPE}
%%%%%%%%%%%%%%%%%%%%%%%%%%%%%%%%%%%FIGURE S01
\begin{figure*}[h!]
\centering
\includegraphics[width=\textwidth]{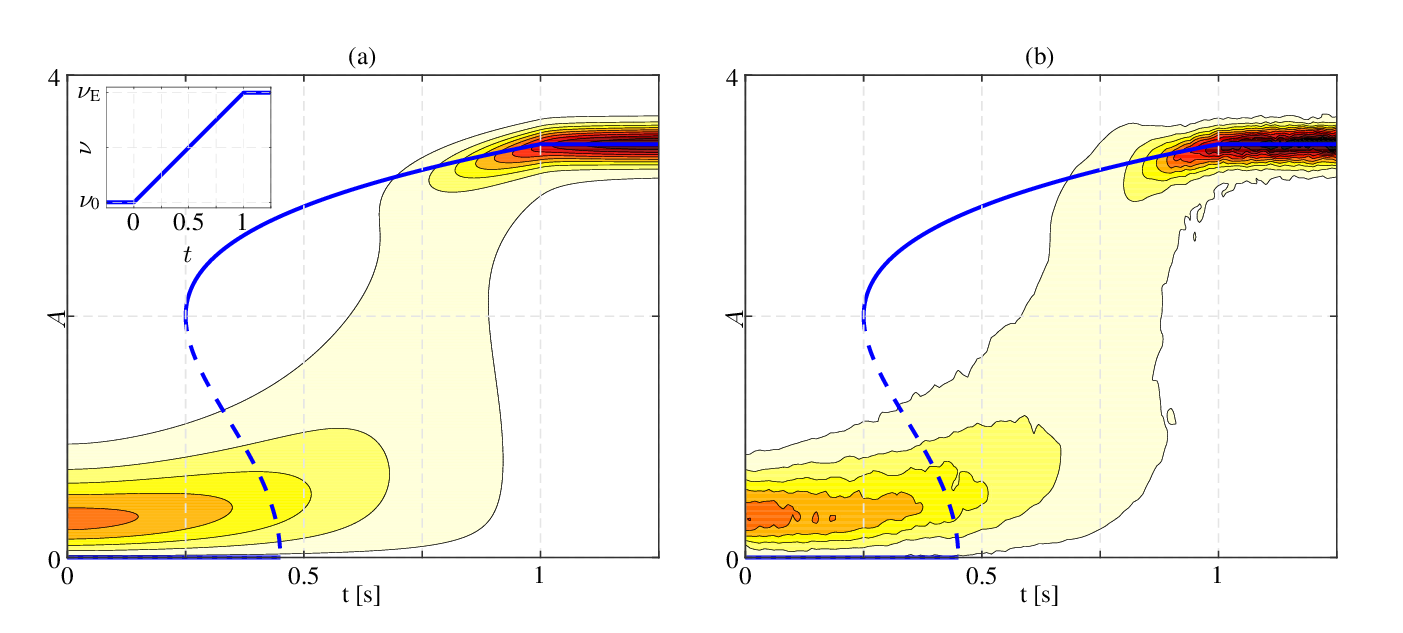}
\caption{Validation of the numerical solution of the FPE (a) with time domain simulation (b) for a ramp rate of $R$=10 rad/s\textsuperscript{2}. The contour plot represents the PDF $P(A;\,t)$ of the ramp process. The stationary deterministic bifurcation diagram under quasi-steady assumption is shown, in blue, as a reference.}
\label{fig:S01_FPE_validation}
\end{figure*}
The evolution in time of the probability density function $P(A ; t)$ during the ramping of the control parameter $\nu$ can be obtained solving numerically \eqref{eq:fp}. In this case, the drift coefficient is time dependent $\mathcal{F}(A;t)=A(\nu(t)+{\kappa A^2}/{8}-{\mu A^4}/{16})+{\Gamma}/{4\omega_{0}^{2}A}$, where $\nu(t)=\nu_0+Rt$. The solution is obtained via the MATLAB\textsuperscript{\textregistered} \verb|ode23| solver, imposing a Dirichlet boundary condition $P(A ; t)$=0, $\forall t$ in $A=0$ and $A=A_\text{max}$, $A_\text{max}$ being the upper boundary of the domain. The initial condition is the stationary PDF $P_\infty(A;\nu_0)$. The result for the set of parameters $\nu_0=-4.5$, $R=10$, $\kappa=8$, $\mu=2$, $\Gamma=10^6$, $\omega_0=120\times2\pi$, $A_\text{max}=6$ is presented in the left panel of \cref{fig:S01_FPE_validation}. The contour plot represents the PDF $P(A ; t)$ for a ramping sequence leading to a significant bifurcation delay. 
This solution of the unsteady FPE is validated against the statistic of Monte Carlo simulations of the process. In detail, \eqref{eq:oscillator} is simulated 5000 times in Simulink\textsuperscript{\textregistered}, imposing again $\nu=\nu(t)=\nu_0+Rt$ and with the initial conditions distributed according to the stationary PDF $P_\infty(A;\nu_0)$. The ensemble statistic of the trajectories are presented in the right panel of \cref{fig:S01_FPE_validation}. Close agreement with the FPE method can be observed.
%
%%%%%%%%%%%%%%%%%%%%%%%%%%%%%%%%%%%%%%%
%%%%%%%%%%%%%%%%%%%%%%%%%%%%%%%%%%%%%%%
\subsection*{Calculation of the First Passage Time distribution using the FPE}
\label{SUP_fpt}
\begin{figure*}
\centering
\includegraphics[trim= 0 0 0 0,clip,width=\textwidth]{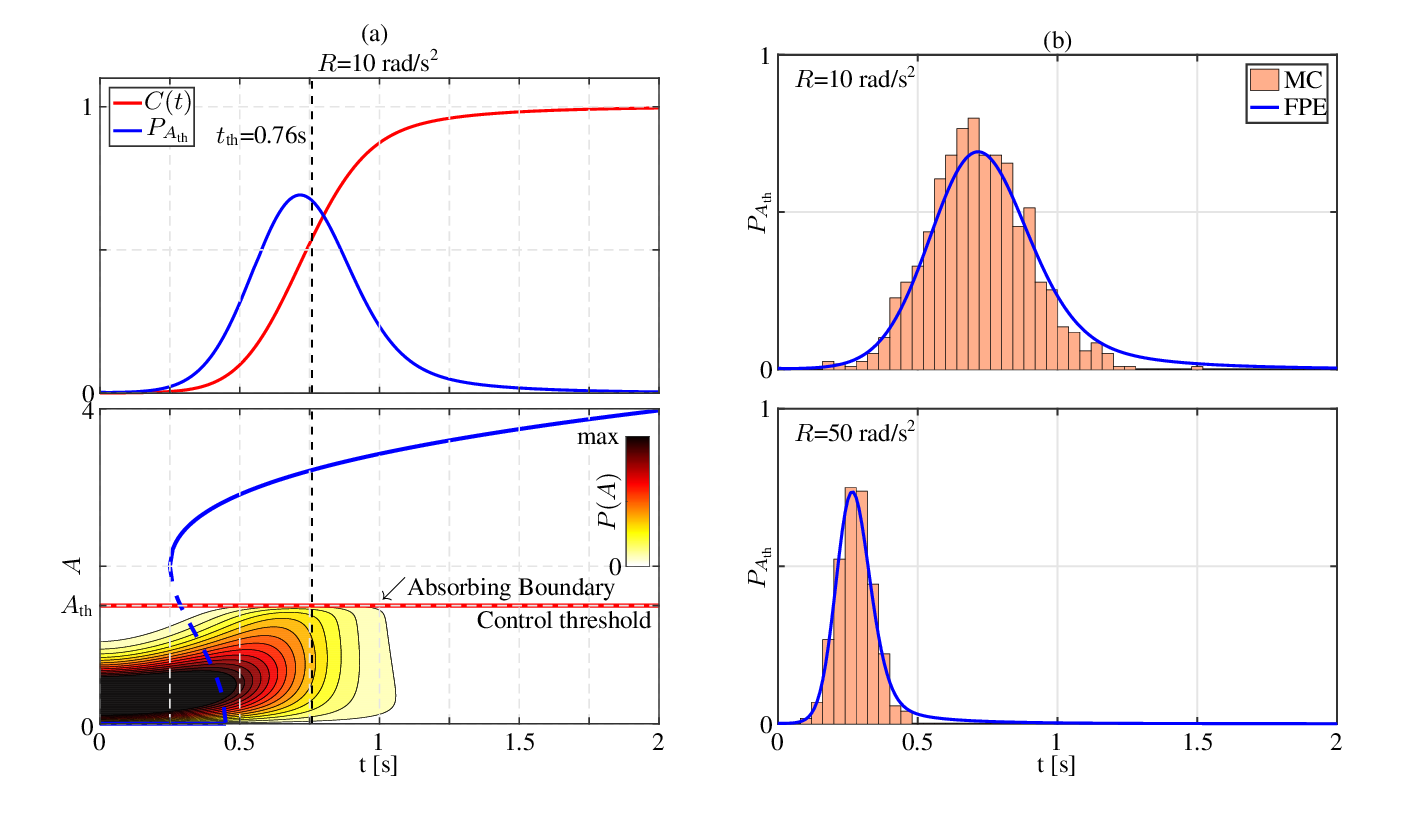}
\caption{a) FPE solution with absorbing boundary condition and related CDF and PDF of first passage time for a ramp rate of R =10 rad/s\textsuperscript{2}. b) Validation of the FPE method against Monte Carlo simulations statistic for two different ramp rates.}
\label{fig:S02_fpt}
\end{figure*}
The following procedure was adopted to compute the distribution of the first passage time $t_\text{th}$ above the threshold $A_\text{th}$. The FPE \eqref{eq:fp}, with $\nu(t)=\nu_0+Rt$ was numerically solved using the MATLAB\textsuperscript{\textregistered} \verb|ode23|, imposing the Dirichlet condition  $P(A=0\,;\,t)=0$ on
the lower boundary and an absorbing boundary condition on the threshold: $P(A\,;\,t>t^*\,|\,A(t*)=A_\text{th}\,;\, t^*)=0$. 
This boundary condition is a probability sink, which leads to a monotonic decay in time of the integral $\int_0^{A_\text{th}}P(A;t)\,\mathrm{d}A$. This integral represents the probability of not having crossed the  threshold $A_\text{th}$ before time $t$. Therefore the probability of having crossed the threshold before $t$ is  the cumulative distribution function (CDF) of the first passage time $C(t)=1-\int_0^{A_\text{th}}P(A;t)\,\mathrm{d}A$. Subsequently the PDF for the first passage time was obtained by differentiating the CDF and the results are given in the left panel of \cref{fig:S02_fpt}. In contrast with the FPE simulation presented in \cref{fig:S01_FPE_validation}, $P(A\,;\,t)$ is soaked up at $A=A_\text{th}$ due to the absorbing boundary condition. In turn, $C(t)$ increases monotonically from zero at $t=0$ and approaches 1 for increasing likelihood of having passed the tipping-point. The PDF $P(t_\text{th})$  (blue curve) is then deduced by differentiating $C(t)$ and  the mean first passage time $\left< t_\text{th}\right>$ is then readily computed as the mean of this probability distribution. In order to validate this FPE-based method, the first passage time probability distribution is computed doing a statistic of 5000 time-domain  simulations of the process in Simulink\textsuperscript{\textregistered}. Similarly to the unbounded case presented in the previous section, simulations are initialised according to the stationary PDF at time 0. For each simulation the first time when the amplitude of the oscillations exceeds the amplitude $A_\text{th}$ is recorded, and the distribution of this time over all the realisations is computed. The results are presented in the right panel of \cref{fig:S02_fpt}: the outcome of the Monte Carlo simulations (histograms) confirms those of the FPE method for all the ramp rates $R$. The map in \cref{fig:06_fpt} was generated repeating the FPE  procedure presented in this section for different $R$ and applying the mapping $\nu_\text{th}=\nu_0+Rt_\text{fp}$. This approach is significantly cheaper from the computational viewpoint compared to the statistic of the time-domain simulations.
%%%%%%%%%%%%%%%%%%%%%%%%%%%%%%%%%%%%%%%
%%%%%%%%%%%%%%%%%%%%%%%%%%%%%%%%%%%%%%%
\subsection*{Supercritical bifurcation}
\label{SUP_supercritical}
%%%%%%%%%%%%%%%%%%%%%%%%%%%%%%%%%%FIGURE S03
\begin{figure*}
\centering
\includegraphics[width=\textwidth]{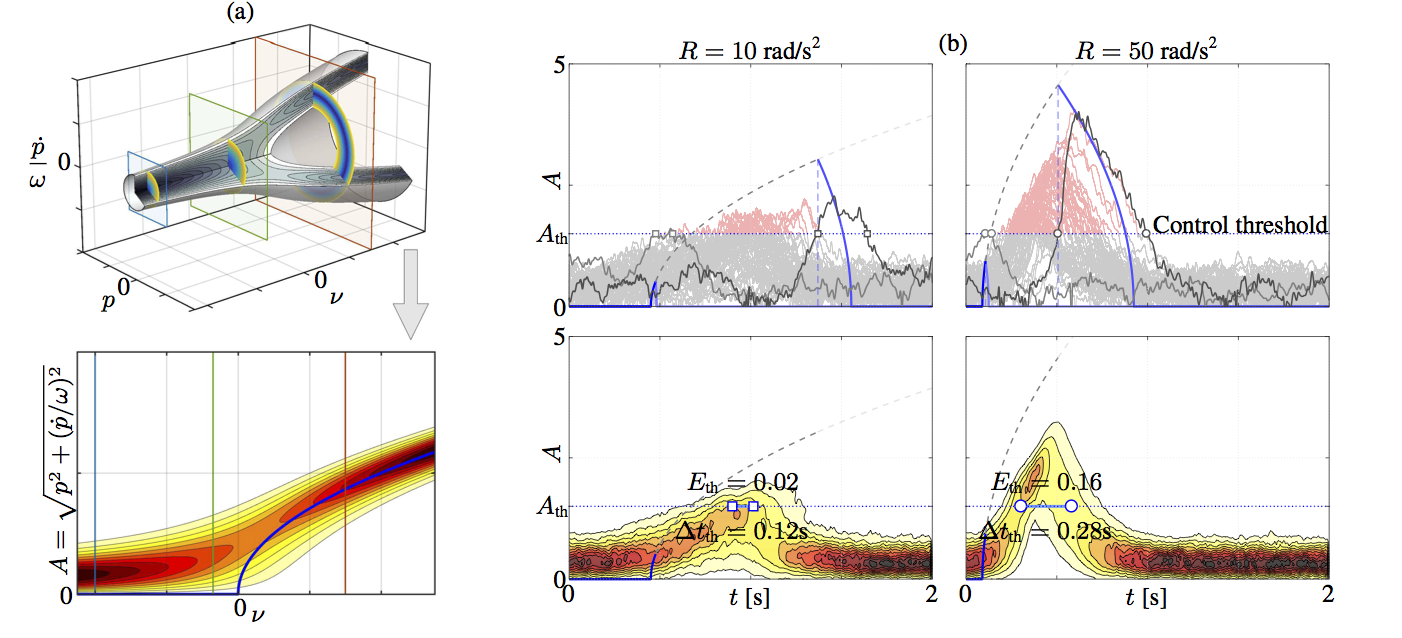}
\caption{a) Stationary PDF $P_\infty(p,\dot{p}/\omega;\nu)$ (top) and stationary amplitude PDF $P_\infty(A;\nu)$ (bottom), for the supercritical bifurcation of the Van der Pol oscillator (\ref{vdp}) (compare with \cref{fig:02_bif}b-c). Adopted parameters values are: $\omega_0/2\pi=120$s\textsuperscript{-1}, $\kappa=8$s\textsuperscript{-1}, $\Gamma/4\omega_0^2=0.44$, $\nu\in[-4.5;5.5]$. b) Same test as in \cref{fig:07_control}, this time applied to the supercritical bifurcation of the Van der Pol oscillator. Two exemplary cases (square $R$=10, circle $R$=50) are simulated in Simulink, implementing a control system that ramps down $\nu$ if the danger level is reached. 
Top row: different realizations of the process (thin lines, gray and red in the safety and danger zones, respectively)
and the two extreme realizations in terms of first passage time (thick gray lines) with their associated quasi-steady deterministic bifurcation diagrams (blue lines). 
Bottom row: KDE of the PDFs. The mean residence and mean released energy in the danger zone are indicated too.}
\label{fig:S03_supercritical}
\end{figure*}
Bifurcation delays exist for any type of bifurcations. In the context of thermoacoustic instabilities in practical combustion chambers, super-critical stochastic Hopf bifurcations are very common. The Van der Pol oscillator with stochastic forcing is the simplest model for this type of bifurcation:
\begin{equation}
\label{vdp}
\ddot{p}+\omega_0^2p=[2\nu -\kappa p^2]\dot{p}+\xi(t),
\end{equation}
with all the terms having the same meaning as in \eqref{eq:oscillator}. The left panel of \cref{fig:S03_supercritical} shows the the probability density function of the amplitude $P_\infty(A\,;\,\nu)$ for a quasi-steady ramping of the bifurcation parameter. As with the surrogate model of the sub-critical Hopf bifurcation, simulations were performed to illustrate the incurred risk when one of the system parameters is ramped at a finite rate while a controller is fed with the current state of the system. The parameter $\nu$  was ramped up in time and as soon as the oscillation amplitude exceeded the control threshold $A_\text{th}$, $\nu$ was ramped back with the maximum possible rate. The results for two different ramp up rates $R$ are shown in the right panel of \cref{fig:S03_supercritical}. Again, it can be observed the averaged released energy is higher when the ramp rate is faster.
\end{document}